\documentclass[twocolumn,nofootinbib,showpacs,prd]{revtex4}%
\usepackage{amsmath}
\usepackage{amsfonts}
\usepackage{amssymb}
\usepackage{graphicx}

\begin{document}

\title{Scale - Free model for governing universe dynamics}

\author{Orlando Luongo}\email{luongo@na.infn.it}
\address{Dip. di Fisica, Universit\`a di Roma "La Sapienza", Piazzale Aldo Moro 5, I-00185 Roma, Italy.\\
ICRANet and ICRA (International Center of Relativistic Astrophysics
Networks), Piazzale della Repubblica 10, I-65122 Pescara,
Italy.\\Dip. di Scienze Fisiche, Universit\`a di Napoli "Federico
II", Compl. Univ. di Monte S. Angelo, Edificio G, Via Cinthia,
I-80126 - Napoli, Italy.}
\author{Carmine Autieri}\email{autieri@sa.infn.it}
\address{Dip. di Fisica "E. R. Caianiello", Universit\`a di
Salerno, I-84081, Baronissi (Salerno), Italy.}

\begin{abstract}
We investigate the effects of scale-free model on cosmology,
providing, in this way, a statistical background in the framework of
general relativity. In order to discuss properties and time
evolution of some relevant universe dynamical parameters
(cosmographic parameters), such as $H(t)$ (Hubble parameter), $q(t)$
(deceleration parameter), $j(t)$ (jerk parameter) and $s(t)$ (snap
parameter), which are well re-defined in the framework of scale-free
model, we analyze a comparison between WMAP data. Hence the basic
purpose of the work is to consider this statistical interpretation
of mass distribution of universe, in order to have a mass density
$\rho$ dynamics, not inferred from Friedmann equations, via scale
factor $a(t)$. This model, indeed, has been used also to explain a
possible origin and a viable explanation of cosmological constant,
which assumes a statistical interpretation without the presence of
extended theories of gravity; hence the problem of dark energy could
be revisited in the context of a classical probability distribution
of mass, which is, in particular, for the scale-free model,
$P(k)\sim k^{-\gamma}$, with $2<\gamma<3$. The $\Lambda$CDM model
becomes, with these considerations, a consequence of the particular
statistics together with the use of general relativity.
\end{abstract}


\maketitle

\section{Introduction}

We found, in literature, many evidences of a lot of unsolved
problems, which are still open questions in the picture of Einstein
field equations. One of the most important issue of general
relativity (GR) is a complete comprehension and interpretation of
the cosmological constant in accordance with data. In the framework
of Friedmann equations we write equations of motion of $\rho$, and
of pressure $p$, in terms of scale parameter $a(t)$; on the other
hand we do not consider any kind of mass dynamics, induced as an
emergent effect of a statical mass distribution too; this could mean
that we lose information about matter dynamics and distribution, in
sense that we consider it as the static part of a dynamic equation,
where the scale factor $a(t)$ is the true dynamical element. In
order to consider a more direct matter dynamics we focus on $\rho$
of universe and we supposed the validity of the scale-free model
\cite{grande2} into the framework of GR; i.e. to say that could
exist a statistical model which describes the motion of matter and
defines implicitly, in this way, the true dynamics of the scale
factor. This interpretation will allow us to describe a dynamics of
universe, starting only from the basic assumptions that every part
of universe (in the framework of homogeneity and of isotropy) is
characterized by a distribution of gravitons, induced by the simple
presence of mass. Every graviton is linked to each other by a
network or by a family of networks and its existence is provided by
the implicit presence of nodes, characterized by a space-time vector
$x^{\mu}$, in which is possible to have a graviton. This model
appears to be useful for matter distribution of universe, making us
able, also, to consider the universe like a disordered system.

In particular, it is possible to describe the existence of
heterogeneous spectrum of local interaction patterns topologically
linked to the presence of nodes. Following the features done by
other authors \cite{grande}, we try to investigate the consequences
of degree heterogeneity on dynamic properties of networks isolating
a region of universe of correlated networks with a completely random
organization, defined by the distribution $P(k)\sim k^{-\gamma}$,
where $\gamma$ has to be considered as an integer number. This
ansatz shows apparently that dynamics would be limited to a
particular region but it is generalized, simply, assuming, as in the
beginning, a homogeneous and isotropic universe, which allows us to
extend a particular result to the whole system (the universe
evolution). Moreover the possibility to consider an elevate number
of networks, $N$, is the basic idea to consider firstly, because,
although $N<\infty$, it is true, at the same time, that $N$ is a
very large number.

In order to describe dynamics, we need all the quantities
\cite{ullando} describing it. These are the cosmographic parameters
$H(t)$, $q(t)$, $j(t)$ and $s(t)$
\cite{visser1,star1,star2,peebles,novex1,novex2}; respectively the
Hubble parameter, the deceleration parameter, the jerk parameter and
finally the snap parameter; these are defined in principle for all
$t$, or redshifts $z$. Assuming the power law degree distribution of
the form $P(k)\sim k^{-\gamma}$, which is the basic idea of the
model, and supposing no interactions among networks, that is
equivalent to say that the type of interaction described by us, i.e.
mean-field type, we are also able to infer how the scale-free model
also deals with transitions and, so, it could be also fruitful to
explain, for a fixed number of gravitons into account, for what
values of the involved quantities the transition era could start,
that could be a real topic to test the statistical model proposed
\cite{bip}.

It is also possible to characterize the model considering the value
of parameter $\gamma$, which is the most influent parameter of the
model. For example some evidences describe a probability
distribution with $2<\gamma<3$, which for real systems appears to be
a correct choice \cite{grande, grande2}. Of course, indeed,
depending to the value of $\gamma$, we find many different
behaviors; in the present paper we do not matter initially about the
exact value of $\gamma$ in order to describe a particular
behavior\footnote{See for details \cite{grande2,VK81}.}.

For our purposes, following the description of
\cite{grande,grande2}, the first quantity introduced to describe the
matter distribution is $\rho_{k}(t)$, which represents the average
density of particles in nodes of degree $k$ and for all the
variables, it holds
\begin{equation}
\dot{\rho}_{k}(t) = - \rho_{k} + k \rho z_{1} + \Lambda[\rho_{k}],
\end{equation}
which is a first example of Langevin equation for our model; it
suggests that the first two terms on the right side of equation are
due to diffusion in uncorrelated random graphs and the third could
be considered as a reaction kernel that depends only on $\rho_k$.
From this equation we derive the dynamics of a particular choice of
$k$. In order to show the evolution of density $\rho(t)$, the
density which is present in Friedmann equations, we must require
that
\begin{equation}
\rho(t)= \sum_{k} P(k) \rho_{k}(t).
\end{equation}
In the continous limit, the above sum becomes an integrals over the
$k$ space, which has to be troncated by using $k^{*}(t) \sim
z_1/\rho(t)$. Easily, following \cite{grande,grande2}
\begin{equation}\label{kappastar}
\int P(k) \rho_{k}^2 dk \propto \left\{ \begin{array}{cc}
\rho^{\gamma-1} + c  k_c^{2-\gamma} \rho & \ \text{if} \ \ k_c >
k^*,
\\   k_c^{3-\gamma} \rho^2 & \ \text{if} \ \ k_c < k^*.  \end{array}
\right.
\end{equation}

We consider that near a critical point $\mu_{c}\Longleftrightarrow
k_c$, density is small, thus we expect that the mean-field regime
holds when $k^{*}
> k_c$ and $N<\infty$. The corresponding Langevin equation is
\begin{equation}\label{langevin}
\dot{\rho}(t) \simeq a  \rho - b \rho^2 + \sqrt{(a \rho + 2 b \rho^2
)/ N}\eta(t)
\end{equation}
where we considered $a, b, c$ three integration constants and, in
particular
\begin{eqnarray}
k_c \propto N^{1/\omega}.
\end{eqnarray}
If we assume universe as a whole system we must require the possible
presence of noise term; we describe an universe without noise if
\begin{equation}\label{iltedellass}
\rho \gg N^{(\gamma-\omega-3)/(2 \omega)},
\end{equation}
with the $\omega$ parameter which has to be in relation with
$\gamma$ because, if eq. ($\ref{iltedellass}$) holds, we must
require
\begin{equation}\label{uaiuhakj}
\gamma<3\omega+3,
\end{equation}
or equivalently at least $\omega>-(1/3)$; because
\cite{grande,grande2} $\omega$ appears to be the maximum degree of
the network size $N$, it must satisfy $\omega>1$ and so our interval
is always in agreement with this request. The basic idea becomes,
so, to start with the scale-free probability, concerning a discrete
interactions among gravitons and so, after defining the density time
evolution (eq. ($\ref{langevin}$)), considering the cosmographic
evolution of parameters of the universe. All the considerations we
will do, have to respect eqs. ($\ref{iltedellass}$) and
($\ref{uaiuhakj}$). It will be possible to link the universe
description also to critical parameter, as said before, in order to
describe the transition era, and it will be clear in the next
sections.

The article is divided as follows: In the next section we describe,
briefly, all the results involved by the use of GR in the framework
of scale-free model for generic redshift $z$; in particular, we
derive the functional form of the dynamical parameter $a(t)$, first,
which will be the basis for the construction of the cosmographic
parameters, later; their own expressions will be found very easily
for generic redshifts without approximation in this section. In the
third section we compare the results, obtained in the previous one,
in the observable limit $z\ll 1$ with comparison with WMAP data,
trying to link also the expressions found with the definition of
$\Omega_{\Lambda}$, which represents the density of cosmological
constant $\Lambda$, i.e. \emph{dark energy}, and we try to infer the
role of $\Lambda$, just in the statistical way considered. The last
section is devoted to some qualitative considerations, conclusion
and perspectives for future works on this direction.

\section{The Model dynamics}

In this section we infer from scale-free model a specific dynamics
for universe, without a particular limit on $z$, or of time $t$, of
the dynamical quantities involved.

Such a  dynamics, defined from the statistical background is easily
valuable, thanks to the first Friedmann equation, which is an
expression that makes a correspondence between $H(t)$ and $\rho$ as
follows
\begin{equation}\label{1}
H^{2}=\alpha\rho+\beta,
\end{equation}
where we have considered, in principle, to take $k=0$ in the above
expression; and using the positions $\alpha=\frac{8\pi G}{3}$ and
$\beta=\frac{\Lambda c^{2}}{3}$, we can imagine to evaluate $\rho$
and its derivative $\dot{\rho}$ by putting them into eq.
($\ref{langevin}$), having, so, a differential equation for $H(t)$,
in the case $\eta(t)=\delta(t)$, which translates the matter
statistical distribution into an astrophysical differential equation
for $H(t)$, which reads
\begin{eqnarray}\label{3}
\dot{H}=-\left(\frac{p\beta}{2}+\frac{b\beta^{2}}{2\alpha}\right)\frac{1}{H}+\,\,\,\,\,\,\,\,\,\,\,\,\,\,\,\,\,\,\nonumber\\
+\left(\frac{p}{2}+\frac{q\beta}{\alpha}\right)H-\frac{b}{2\alpha}H^{3}+\sqrt{\frac{f(H)}{N}}\delta(t).
\end{eqnarray}
It could give us the correspondent evolution of universe dynamics in
terms of $H(t)$ and $a(t)$; in fact, the solution for $H(t)$ is
writable, in terms of $a(t)$, because of definition
$H(t)=\frac{d}{dt}\log a(t)$. All the discussion is completely
global, in sense that, at this stage, we are not considering any
kind of approximation on $t$, or on $z$, as already mentioned
before; that is the same to say that a complete comprehension of
dynamics is, in this paragraph, involved.

For the sake of simplicity we rewrite eq. ($\ref{3}$) at $t\neq 0$,
making the following positions
\begin{eqnarray}\label{4}
a\equiv \frac{p\beta}{2}+\frac{b\beta^{2}}{2\alpha}\nonumber,
\end{eqnarray}
\begin{eqnarray}
b\equiv\frac{p}{2}+\frac{q\beta}{\alpha},
\end{eqnarray}
\begin{eqnarray}
c\equiv\frac{b}{2\alpha}\nonumber.
\end{eqnarray}
and so we get the simple solution for $H(t)$
\begin{equation}\label{5}
H\left(t-t^{*}\right)=A\sqrt{b+\sqrt{d}\tanh\left\{\sqrt{d}\left(t-t^{*}\right)\right\}},
\end{equation}
where $A=\frac{1}{\sqrt{2c}}$. In the previous expression we
consider the position $d=b^2-4ac$ too and we notice that the
integration constant could be considered as an arbitrary time
$t^{*}$. Depending from $d$ sign we could have different evolutions
of $H(t)$; this sign, indeed, cannot be less than zero: A similar
choice, i.e. $d<0$, shows us a definite time interval in which
$H(t)$ (and so $a(t)$) loses its meaning. Hence, for describing a
variable $a(t)$, for all the time, we must have $d>0$. On the other
hand, this is evident, in order to have a solution, which has no
periodicity, into account. This could, also, be tested, by
considering directly its expression, $d=b^{2}-4ac$ and it gets
\begin{equation}\label{cunzo}
d=\frac{p^{2}}{4}+pq\beta\left(\frac{3}{8\pi G}-2\right),
\end{equation}
which is always positive for all the possible values assumed by $p$
and $q$, if we require that $pq>0$. Our choice is the simplest one:
Because of $p,q$ are ''phenomenological'' quantities, defining the
type of reaction and process, we could assume, without losing
information, $p>0$ and $q>0$. Indeed, it is possible in principle,
that $p<0$ and $q<0$, but we will deny this, comparing also with
WMAP data. Our efforts, on the other side, describe a model, whose
validity is strongly dependent from $p,q$ values, which has to be
chosen by data.\\ From integrating eq. ($\ref{3}$), it is possible
to get, also, a precise value for $H(0)$. The theoretical
expression, that we find, is
\begin{equation}\label{6}
{\frac{1}{\alpha ^2}4H^4 N=p\left(\frac{H^2-\beta }{\alpha
}\right)+2q\left(\frac{H^2-\beta }{\alpha }\right)^2},
\end{equation}
which is the solution of equation $H(0)=\sqrt{\frac{f(H(0))}{N}}$,
where $f\left(H\left(t=0^+\right)\right)=f\left(H(0)\right)$,
together with $H\left(t=0^{-}\right)=0$ condition, which is obvious,
if we consider the positive direction of time. Eq. ($\ref{6}$) gives
two positive solutions, i.e.
\begin{equation}\label{unkjkj}
H(0)_{1}= \frac{\sqrt{\frac{p \alpha }{4 N-2 q}-\frac{4 q \beta }{4
N-2 q}-\frac{\sqrt{p^2 \alpha ^2-16 N p \alpha  \beta +32 N q \beta
^2}}{4 N-2 q}}}{\sqrt{2}}\,,
\end{equation}
\begin{eqnarray}\label{khjkhkj}
H(0)_{2}=\frac{\sqrt{\frac{p \alpha }{4 N-2 q}-\frac{2 q \beta }{2
N- q}+\frac{\sqrt{p^2 \alpha ^2-16 N p \alpha  \beta +32 N q \beta
^2}}{2 (2 N- q)}}}{\sqrt{2}}\,.\,\,\,\,\,
\end{eqnarray}
Both solutions are, in the thermodynamic limit, equivalent and they
would read $H_{0}=0$; in our case, for $N<\infty$, only eq.
($\ref{khjkhkj}$) is relevant, if happens that $p\alpha>2q\beta$.
This condition will represent the correct interval for $p$ and $q$
choice.

Moreover, eq. ($\ref{khjkhkj}$) is also useful because tells us the
value of $t^{*}$ in the expression ($\ref{5}$); it reads
\begin{equation}\label{pertstar}
t^{*}=\tanh^{-1}\left(\frac{b-2cH_{0}^{2}}{\sqrt{d}}\right).
\end{equation}
After evaluating the $H(t)$ dynamics we naturally achieve to find
the expression for scale factor $a(t)$. We get
\begin{equation}\label{7bis}
a(t)=a(0)e^{\frac{G\left(t-t^*\right)}{\sqrt{2cd}}},
\end{equation}
where the function $G\left(t-t^{*}\right)$ is expressed by
\begin{displaymath}\label{7tris}
G\left(t-t^{*}\right)-\tilde{G}=
\end{displaymath}
\begin{eqnarray}
=-\sqrt{b - \sqrt{d}}\tanh^{-1}\left\{\frac{\sqrt{b + \sqrt{d}
\tanh\left\{\sqrt{d} \left(t-t^{*}\right)\right\}}}{\sqrt{b -
\sqrt{d}}}\right\}+\nonumber\\
 +\sqrt{b + \sqrt{d}}\tanh^{-1}\left\{\frac{\sqrt{b +
\sqrt{d} \tanh\left\{\sqrt{d} \left(t-t^{*}\right)\right\}}}{\sqrt{b
+ \sqrt{d}}}\right\}\,\,\,\,\,\,
\end{eqnarray}
with the integration constant $\tilde{G}$ of the form
\begin{eqnarray}\label{7quatris}
\tilde{G}=\sqrt{b-\sqrt{d}}\tanh^{-1}\left\{\frac{\sqrt{b}}{\sqrt{b
- \sqrt{d}}}\right\}\nonumber\\
-\sqrt{b + \sqrt{d}} \tanh^{-1}\left\{\frac{\sqrt{b}}{\sqrt{b +
\sqrt{d}}}\right\}.
\end{eqnarray}

\begin{figure}
\includegraphics[scale=0.7]{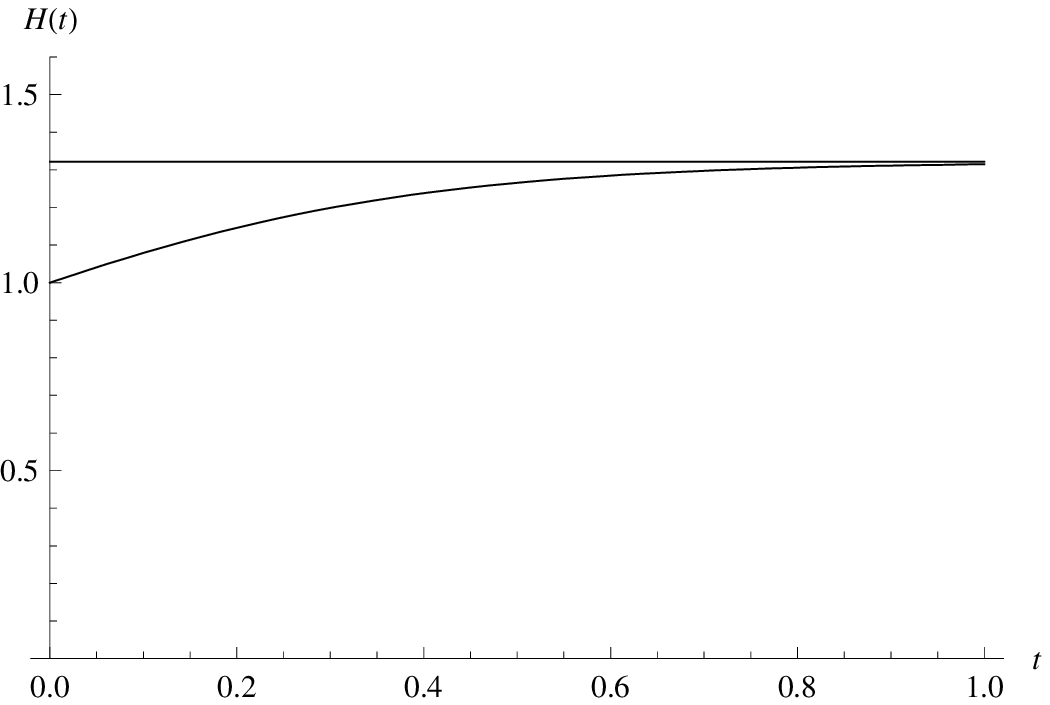}\\
\caption{In this graphic is plotted $H(t)$ with $k=0$. The set of
parameters is $a=1$, $b=3$ and $c=1$, with normalization $H(0)=1$.}
\end{figure}
\begin{figure}
\includegraphics[scale=0.7]{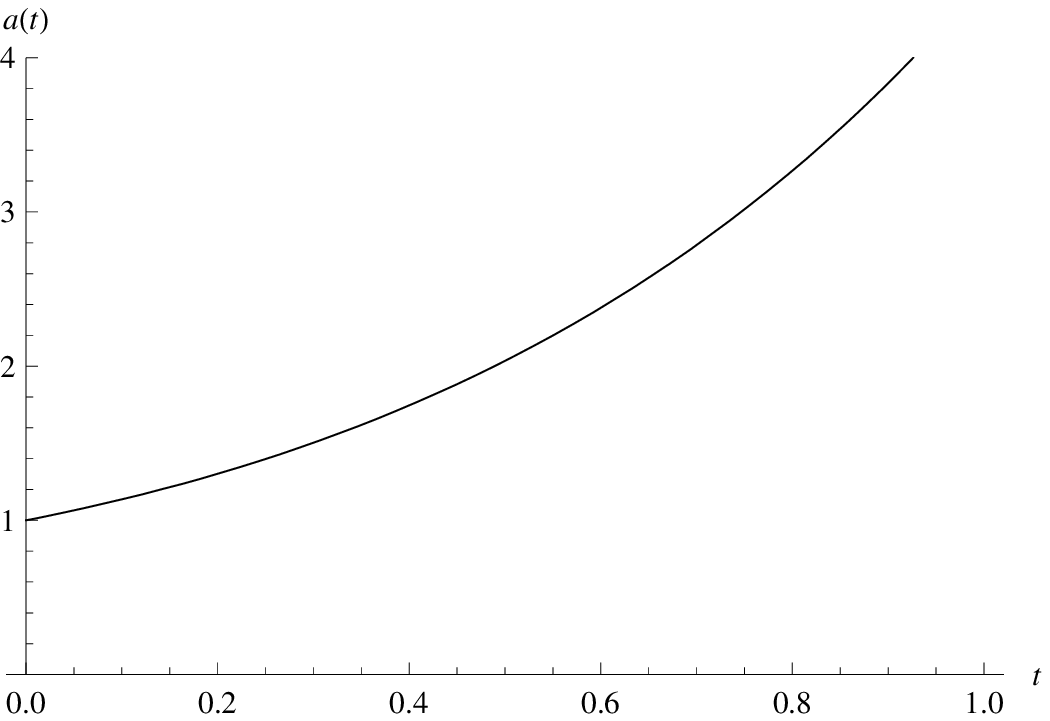}\\
\caption{In this graphic is plotted $a(t)$ with $k=0$. The set of
parameters is $a=1$, $b=3$ and $c=1$, with normalization $a(0)=1$.}
\end{figure}
\begin{figure}
\includegraphics[scale=0.7]{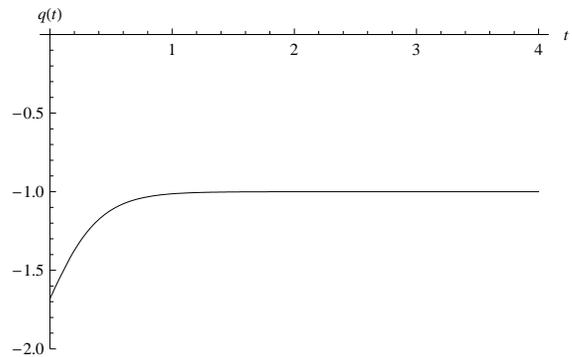}\\
\caption{In this graphic is plotted $q(t)$ with $k=0$. The set of
parameters is $a=1$, $b=3$ and $c=1$.}
\end{figure}

Following all the prescriptions we wrote, and considering the
dynamical scale factor, in order to describe, for all the redshift
$z$, the dynamics of universe, immediately, the expression for the
deceleration parameter, $q(t)$ could be found; it reads
\begin{equation}\label{acceleration}
q(t)=-1-\frac{\sqrt{c/2} d \sqrt{b+\sqrt{d} \tanh\left[\sqrt{d}
\left(t-t^{*}\right)\right]}}{\left(b \cosh\left[\sqrt{d}
\left(t-t^{*}\right)\right]+\sqrt{d} \sinh\left[\sqrt{d}
\left(t-t^{*}\right)\right]\right)^2}.
\end{equation}
It appears, of course, negative, which suggests that the model is
compatible with accelerating universe model. The other expressions
for $j(t)$ and $s(t)$, which discriminates the type of universe
dynamics, are described in appendix for their
complexity.\\

\section{The observable limit}

Recently the possibility to get, from data, all the elements for
universe dynamics, has been reported by several works (see for
example \cite{visser1,Visser2,Visser3,Visser4,star1,star2}). A form
of so-called cosmography, which corresponds to the possibility to
infer, by some useful parameters, the universe evolution, is
involved. We know, in fact, that we are living in the so-called
period of \emph{Precision Cosmology}, in which observations are
extremely precise and allows us to test better than past, a given
evolutionary models. The expressions we need are the so-called
deceleration $q(t)$, jerk $j(t)$, snap $s(t)$
\citep{visser1,star1,star2}. The cosmographic parameters are defined
by derivatives of the scale factor for our epoch; they, instead of
their simplicity, allow to fit the data also in terms of redshift,
that is an implicit measure of time. It is remarkable to remember
that the model, we are going to consider, here and in the following,
is the standard model \cite{Weinberg, visser1, peebles} which is not
useful when $z$ is not little, i.e. $z\ll 1$, that is, indeed, the
case of validity of the Taylor expansion of scale parameter.\\ In
fact, for our time, the cosmographic parameters are defined by the
expression
\begin{eqnarray}\label{tre60aggiungiamotutto}
\frac{a(t)}{a_{0}}=1+H_{0}(t-t_{0})-\frac{1}{2}q_{0}H_{0}^{2}(t-t_{0})^{2}+\nonumber\\
+\frac{1}{6}j_{0}H_{0}^{3}(t-t_{0})^{3}+\frac{1}{24}s_{0}H_{0}^{4}(t-t_{0})^{4}+\ldots,
\end{eqnarray}
and so they reads
\begin{equation}
H(t)\, =\, \frac{\dot{a}}{a},
\end{equation}

\begin{equation}
q(t)=-\frac{\ddot{a}a}{\dot{a}^{2}},
\end{equation}

\begin{equation}\label{jerk}
j(t)=\frac{a^{(3)}a^{2}}{\dot{a}^{3}},
\end{equation}

\begin{equation}\label{snap}
s(t)=-\frac{a^{(4)}a^{3}}{\dot{a}^{4}}.
\end{equation}
We denote with a subscript $0$ these parameters, in sense that they
are evaluated at $t=t_0$; on the meaning of these parameters see for
example \cite{cosmography}. In this paper we will take into account
WMAP observations and so we will assume that the value of the Hubble
constant is $H_0 \simeq 70 \pm 2$ km/sec/Mpc at our time $t=t_{{0}}$
\cite{peebles,wmap}.

Following these advice, it is clear that, the cosmographic
parameters may also be tested by WMAP data \cite{wmap} and, above
all, for the observed limit $z\ll 1$. It is also possible to express
them, in terms of the dark energy density $\Omega_{\Lambda}$, which
could stress also the role of a cosmological constant involved into
Einstein field equations, and so, give a description of $\Lambda$CDM
model in terms of statistical background.

In order to understand such description, we can follow the
prescription of \cite{DETF}. Hence, we will use the
Chevallier\,-\,Polarski\,-\,Linder (CPL) parameterization for the
equation of state setting \cite{CPL}, which reads $w = w_{0} +
w_{a}- \frac{w_a}{1 + z}$, with $a(z)=\frac{a_0}{1+z}$. Easily,
noting the correspondence $t\longleftrightarrow z$, i.e.
 $dt=-\frac{d\log(1+z)}{H(z)}$. Moreover, for
the scenario, we are considering, in the $\Lambda$CDM case, i.e.
\begin{equation}\label{scenario}
\Omega_m+\Omega_k+\Omega_\Lambda=1,
\end{equation}
where $\Omega_m$ is normalized\footnote{Normalized by
$\rho_c\equiv\frac{3H_0^2}{8\pi G}$.} matter density, $\Omega_k$
spatial curvature density and $\Omega_\Lambda$ cosmological constant
density, the list of above parameters has to be set $(w_{0}, w_{a})
= (-1, 0)$, and, then, after some straightforward calculations, it
is useful to note $q_0  =  \frac{1}{2} -
\frac{3}{2}\Omega_{\Lambda}$, $j_0$ and $s_0  =  1 -
\frac{9}{2}\Omega_m$.

Now we have all the ingredients to understand a complete comparison
with scale-free quantities and cosmological parameters. In the next
section, so, will be possible to write down an expression for
$\Lambda$, by comparing with the above expression for
$\Omega_\Lambda$ in terms of $q_0$.

\subsection{The comparison with WMAP data}

After comparing with the experimental value \cite{cimmi-b1,cimmi-b2}
defined by $H_{0},q_{0},s_{0},l_0$, in a easy way, we link
measurable parameters, in terms of scale-free parameters
\begin{equation}\label{15}
q_{0}=-1-\frac{d}{2bH_{0}},
\end{equation}
\begin{equation}\label{16}
j_{0}=1+3\left(\frac{d}{2bH_{0}}\right)-\left(\frac{d}{2bH_{0}}\right)^{2},
\end{equation}
\begin{equation}\label{16bisxxx}
s_{0}=1+\left(\frac{d}{2bH_{0}}\right)\left[6-\left(\frac{d}{2bH_{0}}\right)+3\left(\frac{d}{2bH_{0}}\right)^{2}-\frac{4cd}{b}\right],
\end{equation}
and inverting, we get expression for the scale-free parameters in
function of deceleration, jerk and Hubble parameters as follows
\begin{equation}\label{21}
c=\frac{1-s_{0}-(1+q_{0})\left[7+q_{0}+3(1+q_{0})^{2}\right]}{8H_{0}(1+q_{0})^{2}},
\end{equation}
\begin{equation}\label{22}
b=H_{0}\frac{1-s_{0}-(1+q_{0})\left[7+q_{0}+3(1+q_{0})^{2}\right]}{4(1+q_{0})^{2}},
\end{equation}
\begin{equation}\label{23}
d=-2H_{0}^{2}\frac{1-s_{0}-(1+q_{0})\left[7+q_{0}+3(1+q_{0})^{2}\right]}{4(1+q_{0})}.
\end{equation}
All the previous results deal with a dynamics, induced by the use of
the scale-free model, and treat the link among scale-free parameters
and the observed parameters.\\ It is necessary to note that all the
expressions obtained are in the case $k=0$, which is the observed
one \cite{peebles}. Inducing a curvature term $k\neq 0$ is possible
to achieve an expression for $a(t)$, which is not analytically
defined. It appears as a strong complication of the model; indeed,
it seems so also because observation provide a curvature density
$\Omega_k\sim 0$, but we could easily consider a graphic of it, in
order to fix ideas.
\begin{figure}[h]
  \includegraphics[scale=0.7]{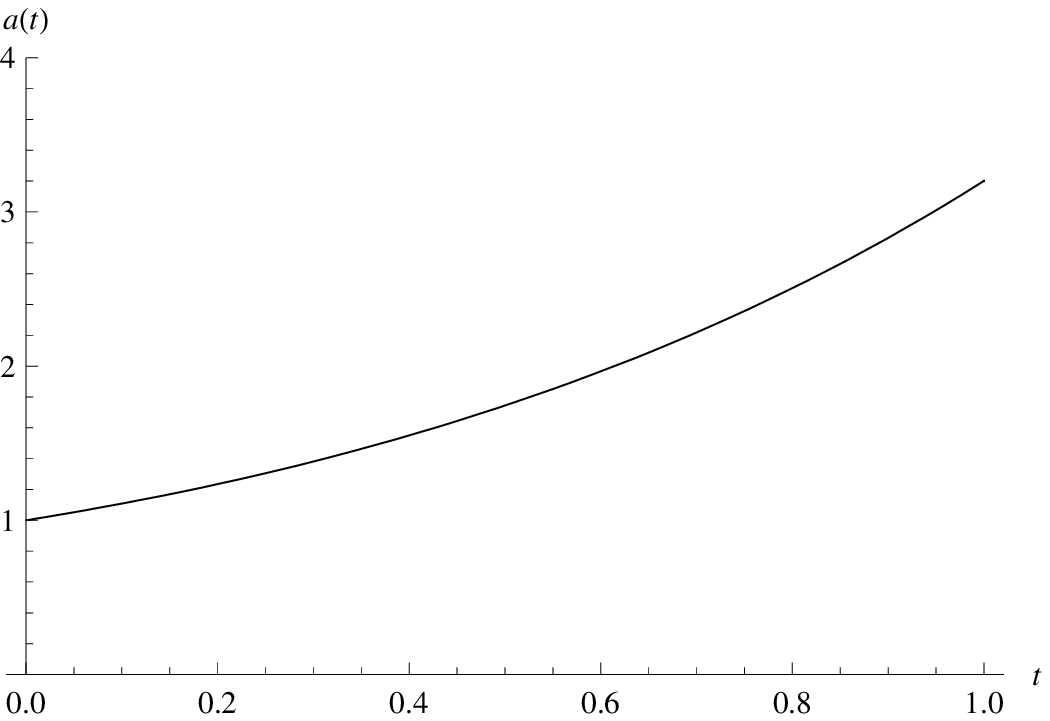}\\
  \caption{In this graphic is plotted $a(t)$ with $k\neq0$. The set of
parameters is $a=1$, $b=3$ and $c=1$, with normalization $a(0)=1$.}
\end{figure}
It is important for us to underline that all the graphics are
evaluated by a specific choice of parameters $a,b,c$ reported in the
graphic labels, not chosen by particular reasons but only for
graphic simplicity.

Now we can write down all the parameters in terms of $H_{0}$, as
follows
\begin{eqnarray}\label{21}
c\sim 0.35H_{0}^{-1}\,,\nonumber\\
b\sim -2H_{0}\,,\,\,\,\,\,\,\,\\
d\sim 2H_{0}^{2}\,.\,\,\,\,\,\,\,\,\,\,\,\nonumber
\end{eqnarray}
and eqs. ($\ref{bleeding}$ and $\ref{21}$) allow us to write down
cosmographic parameters in terms of $\Omega_{\Lambda}$ and, in
particular, we find the definition of $\Lambda$ in our model as
follows
\begin{equation}\label{lambdax}
\Lambda\sim\frac{7\rho_c}{\cosh^2\left(\sqrt{2}H_0 t_0\right)}+1.
\end{equation}

Remembering the definition of scale-free parameters, postulated at
the beginning, in particular of critical transition parameter, we
find for it
\begin{equation}\label{pix}
k_c\sim1.429\,H^{3}_0.
\end{equation}
In other words the effect of a dark energy are modeled by the
presence of a cut off $k_{c}$ and for our epoch, in which $k>k_{c}$,
all the previous results suggest how to consider the correct meaning
of $\Lambda$: It could be considered as an effect of modeling
universe with a $P(k)\sim k^{-\gamma}$, or, in other words, as a
statistical result.

\section{Conclusion and perspectives}

We investigated in this paper that the scale-free model, achieving a
statistical interpretation of the ''background'' of space time, by
the presence of a large number of gravitons, allowed us, thanks to
networks interactions, to define directly $\rho$ dynamics, without
using, the scale factor $a(t)$ dynamics, in the picture of mean
field approximation.

Moreover we showed, that, if that dynamics is involved into the
framework of GR, we are able to define all the cosmographic
parameters, in terms of time evolution and, in principle, also of
redshift $z$. Best results were $q(t)<0$ which represents the
acceleration parameter of universe and tells us the universe
expansion and also the cosmological constant $\Lambda$, which became
dependent from statistical and cosmological parameters. This could
be thought as an explanation of the peculiar reason for what a
cosmological constant has to be put into Einstein field equations,
in order to respect the correct statistical matter distribution and
matter evolution.

We have, also, discussed the possibility to evaluate the parameters
of scale-free theory at our epoch in terms of measured cosmographic
parameters and vice-versa.

Starting from these results, we can discuss, indeed, also the
possibility of defining a dark matter dynamical equations thanks to
the expression of density motion, defined in the introduction and,
of course, we can also focus a possible reformulation of $\Lambda$
in the framework of a microscopic theory, involved by the scale-free
model directly.

The choice of $P(k)\sim k^{-\gamma}$ appears to be comparable with
the density of a high-energy astrophysical object and could suggest
that scale-free model could be a possible adaptable statistical
model for the cases of Fermi processes \cite{peebles}.

The scale-free model appears to be a framework model for gravitation
so, in principle, extensions of GR are adaptable with scale-free
model.

Moreover, all the previous results, concerning finite networks,
could be extended in the case of infinite networks by using
diverging cut-off $k_c$. It is possible to show, that depending from
$\gamma-1$ and $q$ values, different results are involved, and this
will concern to future efforts.

Other fields of future interest are to be found in the possibility
to join the statistical behavior, described in this work, not
necessary in the case of GR \cite{my33}. It will be clear in future
work, that modifications occur if a different field theory is
involved; idea which is possible, because the model appears to be
independent from the field background. Concluding, of course, the
statistics in this sense is a first kind of example in which the
possibility to consider a different approach to the problem of
$\Lambda$. This is possible, also passing through a quantum picture
as showed, for example, in \cite{entanglement1}. The deep difference
of this kind of work is the classical way in which all the
calculations have been evaluated.

\section*{Acknowledgements}

The authors warmly thank  prof. S. Capozziello, prof. L. Dall'Asta
and dr. F. Caccioli for important and deep discussions.

\appendix

\section{Cosmographic parameters}\label{sA}


The jerk parameter $j(t)$ is defined as follows
\begin{eqnarray}\label{jerk}
\frac{j(t)}{f(t)}=\beta_1+\beta_2\cosh4\sqrt{d}t+4\cosh2\sqrt{d}t\left(\beta_3+g_4\right)+\nonumber\\
+\beta_5\sinh(4\sqrt{d}t)+g_6+
2\sqrt{d}\sinh(2\sqrt{d}t)\left(\beta_7+g_8\right),\,\,\,\,\,\,\,\nonumber
\end{eqnarray}
where
$f(t)=\frac{\cosh^{-4}\left(\sqrt{d}t\right)}{8\sqrt{2}}\left[b+\sqrt{d}\tanh\left(\sqrt{d}t\right)\right]^3$,
and $g_{n}\equiv \frac{\beta_n H(t)}{A}$. The snap parameter $s(t)$
is
\begin{eqnarray}\label{snap2}
\frac{s(t)}{h(t)}=g_{11}-g_{12}+g_{13}+g_{14}+
\cosh\left(6\sqrt{d}t\right)(g_{15}+g_{16}+g_{17})+\nonumber\\
\sinh\left(2\sqrt{d}t\right)(\beta_9+g_{18}-g_{19}-g_{20}+g_{21})+\,\,\,\,\,\,\,\,\,\,\,\,\,\,\,\,\,\,\,\,\,\,\,\,\,\,\,\,\,\,\,\,\nonumber\\
+\sinh\left(4\sqrt{d}t\right)(\beta_{10}+g_{22}-g_{23}-g_{24})+\,\,\,\,\,\,\,\,\,\,\,\,\,\,\,\,\,\,\,\,\,\,\,\,\,\,\,\,\,\,\,\,\nonumber\\
\,\,\,\,\,\,\,\,\,\,\,\,\,\,\,\,\,\,\nonumber\\
+\sinh\left(6\sqrt{d}t\right)(g_{25}+g_{26})+\,\,\,\,\,\,\,\,\,\,\,\,\,\,\,\,\,\,\,\,\,\,\,\,\,\,\,\,\,\,\,\,\,\,\,\,\,\,\,\,\,\,\,\,\,\,\,\,\,\,\nonumber\\
+2\cosh\left(4\sqrt{d}t\right)(\beta_{27}+g_{28}+g_{29}-g_{30}-g_{31})+\,\,\,\,\,\,\,\,\,\,\,\,\,\,\,\,\,\,\,\,\,\nonumber\\
-\cosh\left(2\sqrt{d}t\right)(\beta_{32}-g_{33}+g_{34}+g_{35}+g_{36}).\,\,\,\,\,\,\,\,\,\,\,\,\,\,\,\,\,\,\,\,\,\,\,\,\,\,\,\,\,\,\,\,\,\,\,\,\,\,\nonumber
\end{eqnarray}
The complete list of constants involved into calculations is
\begin{eqnarray}
\beta_1=\sqrt{2}(3b^3-3bd+4cd^2),
\beta_2=\sqrt{2}b(b^2+3d),\nonumber
\end{eqnarray}
\begin{eqnarray}
\beta_3=\sqrt{2}(b^3-2cd^2), \beta_4=3\sqrt{c}bd,
\beta_5=\sqrt{2d}(3b^2+d),\nonumber
\end{eqnarray}
\begin{eqnarray}
\beta_6=12b\sqrt{c}d,\beta_7=-\sqrt{2}(-3b^2+d+4bcd),\beta_8=6\sqrt{c}d,\nonumber
\end{eqnarray}
\begin{eqnarray}
\beta_9=-8\sqrt{2c}d^{3/2}(-9b^2+3d+10bcd),
\end{eqnarray}
\begin{eqnarray}
\beta_{10}=4\sqrt{2c}d^{3/2}(9b^2+3d+8bcd),\beta_{11}=10b^4,\beta_{12}=-12b^2
d,\nonumber
\end{eqnarray}
\begin{eqnarray}
\beta_{13}=2d^2, \beta_{14}=56bcd^2,\beta_{15}=b^4,\beta_{16}=6b^2
d, \beta_{17}=d^2,\nonumber
\end{eqnarray}
\begin{eqnarray}
\beta_{18}=20b^3\sqrt{d}, \beta_{19}=-12bd^{3/2}, \beta_{20}=-64b^2
cd^{3/2}, \nonumber
\end{eqnarray}
\begin{eqnarray}
\beta_{21}=56cd^{5/2}, \beta_{22}=16b^3 \sqrt{d}, \beta_{23}=-32b^2
cd^{3/2},\nonumber
\end{eqnarray}
\begin{eqnarray}
\beta_{24}=-32cd^{5/2},
\beta_{25}=4b^3\sqrt{d},\beta_{26}=4bd^{3/2},\nonumber
\end{eqnarray}
\begin{eqnarray}
\beta_{27}=2\sqrt{2c}d(3b^3+b(9+4bc)d+4cd^2),\beta_{28}=3b^4,\nonumber
\end{eqnarray}
\begin{eqnarray}
\beta_{29}=6b^2 d, \beta_{30}=-d^2, \beta_{31}=-32bcd^2,\nonumber
\end{eqnarray}
\begin{eqnarray}
\beta_{32}=16\sqrt{2c}d(-3b^3+2b^2cd+3cd^2),
\beta_{33}=-15b^4,\nonumber
\end{eqnarray}
\begin{eqnarray}
\beta_{34}=6b^2 d, \beta_{35}=d^2, \beta_{36}=8bcd^2\nonumber.
\end{eqnarray}

\end{document}